%% file: lyb.tex
\documentclass[11pt,a4paper]{article}
\pdfoutput=1
\usepackage{jcappub}
\usepackage{lineno}
\usepackage{epstopdf}
\usepackage{subfigure}
\usepackage{rotating}
\usepackage{float}

\input{mydefinitions.tex}

\newcommand\ion[2]{#1$\;${\small \uppercase\expandafter{\romannumeral #2}}}%
\newcommand\ionalt[2]{#1$\;${\scriptsize \uppercase\expandafter{\romannumeral #2}}}%

\renewcommand{\lya}{Ly$\alpha$}
\newcommand{\lyb}{Ly$\beta$}

\newcommand{\Si}{SiIII\ }
\newcommand{\Ox}{OVI\ }

\graphicspath{{figures/}}

\title{Detection of \lyb\ auto-correlations and \lya-\lyb\ cross-correlations in BOSS Data Release 9}

\input{authors_fix.tex}
\emailAdd{vid.irsic@fmf.uni-lj.si}
\emailAdd{anze@bnl.gov}

\abstract{ The Lyman-$\beta$ forest refers to a region in the spectra
  of distant quasars that lies between the rest-frame Lyman-$\beta$
  and Lyman-$\gamma$ emissions. The forest in this region is dominated
  by a combination of absorption due to resonant \lya\ and \lyb\
  scattering.  When considering the 1D \lyb\ forest in addition to the
  1D \lya\ forest, the full statistical description of the data requires
  four 1D power spectra: \lya\ and \lyb\ auto-power spectra and the
  \lya-\lyb\ real and imaginary cross-power spectra. We describe how
  these can be measured using an optimal quadratic estimator that
  naturally disentangles \lya\ and \lyb\ contributions. Using a sample
  of approximately 60,000 quasar sight-lines from the BOSS Data
  Release 9, we make the measurement of the
  one-dimensional power spectrum of fluctuations due to the \lyb\
  resonant scattering.  While we have not corrected our measurements for
  resolution damping of the power and other systematic effects carefully 
enough to use them for cosmological
  constraints, we can robustly conclude the following: i) \lyb\ power
  spectrum and \lya-\lyb\ cross spectra are detected with high
  statistical significance; ii) the cross-correlation coefficient is
  $\approx 1$ on large scales; iii) the \lyb\ measurements are
  contaminated by the associated \Ox absorption, which is analogous to
  the \Si contamination of the \lya\ forest.  Measurements of the
  \lyb\ forest will allow extension of the usable path-length for the \lya\
  measurements while allowing a better understanding of the
  physics of intergalactic medium and thus more robust cosmological
  constraints.

 }


\begin{document}
\maketitle

\section{Introduction}

Lyman-$\alpha$ forest is a series of absorption lines, blue-ward of
the \lya\ emission in the spectra of high-redshift quasars.  Although
it was discovered nearly half a century ago
\cite{1971ApJ...164L..73L}, it only recently became a useful
cosmological probe
\cite{2003AIPC..666..157W,2003dmci.confE..18W}. An important part in
this evolution was due to the technological progress that made it
possible for large surveys such as Sloan Digital Sky Survey (SDSS,
\cite{2000AJ....120.1579Y,2005ApJ...633..560E,2013arXiv1303.4666A,2013AJ....145...10D,2012AJ....144..144B,2006AJ....131.2332G,1998AJ....116.3040G,1996AJ....111.1748F,2012arXiv1208.2233S,2011AJ....142...72E,2013A&A...552A..96B,2012ApJS..203...21A,2013JCAP...03..024K,2012ApJS..199....3R,2011ApJ...729..141B,2013AJ....145...69L,2012A&A...548A..66P,2012A&A...547L...1N})
to measure spectra of quasars reliably and in large numbers.

The physical picture of the \lya\ forest was established in the 1990s.
The absorption features primarily arise in the near-mean density
regions
\cite{1994ApJ...437L..83B,1994ApJ...437L...9C,1995AAS...187.8503Z}
from the weakly non-linear fluctuations of gas held in equilibrium by
photo-ionizing background radiation
\cite{1965ApJ...142.1633G,1994SPIE.2198..362V}. This makes it possible
for the \lya\ forest fluctuations to be predicted from first
principles using large numerical simulations. Namely, the complicated
astrophysics of fluid dynamics, baryon-condensation, star-formation
and feedback due to supernova and active galactic nuclei activity is
absent - a typical line of sight does not pierce through a virialized
object, and when it does, it results in a complete absorption which
makes the detailed modeling of virialized regions inessential.
Even though, the effect of astrophysics cannot be completely neglected
and has some impact on flux statistics
(\cite{2005ApJ...635..761M,2013MNRAS.429.1734V}), the effect for \lya\
absorption is small and can be largely neglected for quantitative
studies. In particular \cite{2013MNRAS.429.1734V} shows that the
effect of feedback from active galactic nuclei and supernovae falls
off rapidly towards higher redshifts at which our measurements are
taken and is of the order of a percent.  While this makes predictions
of \lya\ quantities considerably easier than \emph{a-priori} galaxy
evolution, the physics of intergalactic medium (IGM) remains
complicated and any results must be cross-checked in as many different
ways as possible.

The field has settled on using the one-dimensional power spectrum
$P_F(k,z)$ of the relative fluctuations in the transmitted flux
fraction $\delta_F$ as the quantity of choice when comparing
observations with the theoretical
predictions\cite{1998ApJ...495...44C,2000ApJ...543....1M,2002ApJ...581...20C,2003ApJ...585...34M,2004MNRAS.355L..23V,2005ApJ...635..761M}. The
main reason for this selection is that the power spectrum of transmitted flux
fluctuations is observationally closest to the data: it is essentially
an appropriately scaled version of the actual fluctuations in the
observed forest and hence it is easy to understand
the systematics and the noise properties of the measurement. Choosing the power
spectrum over the correlation function more cleanly decouples the scales
involved. For example, fluctuations due to poor understanding of the
continuum are restricted to large scales.

Recently, the three-dimensional correlations have been measured in the \lya\
forest
(\cite{2011JCAP...09..001S,2013A&A...552A..96B,2013JCAP...04..026S}) 
and it may eventually be possible to make a
unified analysis of both 1D and 3D correlations. However, systematic
issues are very different in the two cases and at present the 1D
power spectrum of fluctuations is our best approach for measuring the
linear power spectrum amplitude at scales around $k\sim 1h/{\rm Mpc}$.

As discussed above, systematic control of astrophysical and
instrumental effects remains one of the largest challenges in the
\lya\ 
studies. There are two main ways to independently measure the
properties of the IGM and thus cross-check the assumptions. The first
one is to use a higher order statistic (bispectrum or trispectrum,
\cite{2005ApJ...635..761M,2012arXiv1202.3577G,2003MNRAS.344..776M}). This
approach allows one to measure essentially the same quantities as in the
power spectrum with a similar signal-to-noise, but with largely
independent or differently-scaling systematics. An alternative is to
use higher order Lyman absorption, which was proposed in \cite{2004ApJ...605....7D}
and  which we study in this work.

Understanding the \lyb\ forest would be useful in several ways. First and
foremost, the \lyb\ forest probes the same hydrogen gas, but with a
smaller optical depth at a given column density of
gas. Fortunately, there is no uncertainty in the ratio of optical
depths, since it  is entirely determined by atomic physics.
The ratio of cross sections for Lyman series lines simplifies to the
ratio of oscillator strength for those lines, which can be calculated
analytically. The oscillator strength of Lyman transition of order $n$
is given by (\cite{2011aas..book.....P})
\begin{equation}
f_n = 2^8 n^5 \frac{\left( n-1\right)^{2n-4}}{3\left(n+1\right)^{2n+4}}.
\end{equation}
The ratio of the optical depths for $\beta$ and $\alpha$ lines
$r_{\beta\alpha}$ is thus given by 
\begin{equation}
 r_{\beta\alpha}=\frac{\tau_\beta}{\tau_\alpha} = \frac{f_3}{f_2} \approx 0.1901.
\label{eq:3}
\end{equation}
Given that $r_{\beta\alpha}$ is of $O(1)$ (rather than $\ll 1$) means
that we are probing somewhat larger gas densities, but that the
dominant physics is the same and the numerical simulations made for
\lya\ will likely suffice.  While virialized regions are still going
to result in complete absorption, the \lyb\ forest will likely be
affected more by the effects of the galactic feedback (although this
will need to be checked using numerical simulations in further work)
and other nuances of the IGM physics.  Therefore, when used in
conjunction with the \lya\ absorption, \lyb\ information can break
degeneracies in modeling of these regions.

At the same time, the absorption in the \lyb\ region of the forest is
dominated by the \lya\ absorption. Therefore, if one is able to
simultaneously model the \lya\ and \lyb\ regions, it is possible to
extend the useful path length for \lya\ forest by up to 20\% (depending
on the redshift distribution of quasars in a given survey). This can,
for example, significantly increase the sensitivity to the baryon acoustic
oscillations signal, without any increase in the cost of an
experiment.

The purpose of this paper is to make a proof-of-concept measurement of
the \lyb\ forest in the DR9 data release of Baryon Oscillation
Spectroscopic Survey (BOSS;
\cite{2013AJ....145...10D,2012AJ....144..144B}), 
which is part of the Sloan Digital Sky
Survey III collaboration (\cite{2011AJ....142...72E,2006AJ....131.2332G,1998AJ....116.3040G,2000AJ....120.1579Y,1996AJ....111.1748F,2012arXiv1208.2233S}). 
We believe our detection significance is
robust and the results are correct and consistent
with expectations. However, these measurements should not be used to
constrain cosmological parameters: our understanding of the resolution
uncertainty, noise bias and other subtleties is limited. 
Moreover, the results are strong enough to show that these
measurements are clearly feasible with high precision. For example, even
with our limited understanding of systematics, we are able to measure
a contaminating metal line in the \lyb\ forest with percent level
accuracy on its wavelength and identify it as the \Ox feature.

The paper is structured as follows.  In Section \ref{sec:2} we present the
theoretical description of the fluctuations and how physically
relevant quantities can be derived from the data. The data and
simulations used are discussed on in Section \ref{sec:data}. In
Section \ref{sec:mock} we present the results on the mock data and in
Section \ref{sec:result} we show the final measurements on the
data. We conclude in Section \ref{sec:conc}.

\section{Description of the \lya\ and \lyb\ forests}
\label{sec:2}

\subsection{Power spectra of fluctuations}

The spectrum for a quasar $q$ at an observed wavelength
$\lambda_o$ is given by
\begin{equation}
  f^q(\lambda_o) = C^q(\lambda_r) F^q(\lambda_o),
\end{equation}
where $C^q(\lambda_r)$ is the intrinsic quasar spectrum (observed by
an observer in the rest frame of the quasar with redshift $z_q$,
where $\lambda_r=\lambda_o/(1+z_q)$)) and $F(\lambda_o)$ is the total
absorption due to absorbing material along the line of sight to the
quasars
\begin{equation}
  F^q(\lambda_o) = \prod_{i,(z_i<z_q)} e^{-\tau^q_i (r=c \ln
    \lambda_o/\lambda_i)},
\label{eq:Fo}
\end{equation}
where $\tau_i$ is the optical depth for the $i$-th component absorbing
at rest-frame $\lambda_i$ and c is speed of light. Optical depth is a
function of distance, which we parametrise in terms of the logarithm
of the observed wavelength. The reason for this choice is that the
difference in this distance measure is expressed in the usual units of
$\mathrm{km s^{-1}}$. The crucial point is that for a given observed-frame
wavelength, we allow for several absorbers that occupy different
positions along the line of sight to the
quasar. Of course, since matter behind the quasar cannot absorb light,
any given component can absorb only at sufficiently small observed
wavelengths. In other words, \lya\ absorption can be found blue-ward of
the rest-frame \lya\ emission, the \lyb\ absorption blue-ward of the
rest-frame \lyb\ emission, etc.

In \lya\ forest studies, it is usually assumed that the \lya\ absorption
is the dominant source of absorption and worked in terms of the relative
transmitted flux fluctuations 
\begin{equation}
  F^q(\lambda_o) = e^{-\tau^q_\alpha} = \bar{F}_\alpha(r^\alpha)
  (1+\delta^q_\alpha(r^\alpha)),
\end{equation}
where $r^\alpha=c\ln \lambda_o/\lambda_\alpha$ is our radial
coordinate.  We therefore describe the fluctuations in the forest as
relative fluctuations around the mean absorption. The mean of those
fluctuations is $\left< \delta^q \right>=0$ and the two point function
is conveniently described in terms of the correlation function
$\xi_{\alpha\alpha}(x,z)$
\begin{equation}
\left< \delta_\alpha(r^\alpha_1) \delta_\alpha(r^\alpha_2) \right> = 
\xi_{\alpha\alpha}(x=r^\alpha_2-r^\alpha_1=\ln{\lambda_2/\lambda_1}, \bar{z}).
\label{eq:1}
\end{equation}
or equivalently the power spectrum
\begin{equation}
\xi_{\alpha\alpha}(x,{\bar z}) \equiv \frac{1}{2\pi} \int_{-\infty}^\infty
  P_{\alpha\alpha}(k,\bar{z})e^{-kx} dk = \frac{1}{\pi} \int_0^\infty
  P_{\alpha\alpha}(k,\bar{z})\cos(kx) dk,
\label{eq:2}
\end{equation}
where ${\bar z}$ is defined as
\begin{equation}
1 + {\bar z} = \frac{\sqrt{\lambda_1 \lambda_2}}{\lambda_\alpha}.
\end{equation}
Here and henceforth in this paper $\lambda_1$ and $\lambda_2$ are
observed wavelengths ($\lambda_o$) at two different positions in the
quasar spectrum. They should not be confused by rest-frame wavelength
of absorbing material $\lambda_i$ from Equation \ref{eq:Fo} which is,
in this work, replaced by \lya\ rest-frame absorption wavelength
$\lambda_\alpha=1215.67$\AA\ and \lyb\ rest-frame absorption
wavelength $\lambda_\beta=1025.72$\AA.

The power spectrum in Equation \ref{eq:2} is consistent with standard definitions
found elsewhere in the literature 
\cite{2002ApJ...581...20C,2006ApJS..163...80M,2004MNRAS.355L..23V}.

We proceed by  adding the absorption by the \lyb\ line. In this case, where
considering a pixel in the \lyb\ forest, we have
\begin{equation}
    F^q(\lambda_o) = e^{-\tau^q_\alpha-\tau^q_\beta} = 
\bar{F}_\alpha(z_\alpha) \bar{F}_\beta(z_\beta)
  (1+\delta^q_\alpha(r^\alpha))(1+\delta^q_\beta(r^\beta)) =\bar{F}_T(\lambda_o) (1+\delta_T(\lambda_o)).
\end{equation}
Any given pixel in the \lyb\ forest thus receives a contributions to the
absorption from gas residing at two distinct redshifts.  One can
distinguish between the two components only statistically, by
observing the total relative fluctuation $\delta_T$ and
cross-correlating it with other fluctuations in the \lya\ and
\lyb\ forests (see section \ref{sec:meas-power-spectra}). The two-point
function of the \lyb\ forest is given by Equations (\ref{eq:1}) and
(\ref{eq:2}). The cross-power is slightly more subtle:
\begin{equation}
\left< \delta_\alpha(r_1^\alpha) \delta_\beta(r_2^\beta) \right> = 
\xi_{\alpha\beta}\left(x=r^\beta_2-r^\alpha_1=\ln\left[\frac{(\lambda_2/\lambda_\beta)}{(\lambda_1/\lambda_\alpha)}\right],\bar{z}_{\alpha\beta}\right),
\end{equation}
where ${\bar z}_{\alpha\beta}$ is defined as
\begin{equation}
1 + {\bar z}_{\alpha\beta} = 1 + {\bar z}_{\beta\alpha} =
\sqrt{\frac{\lambda_1 \lambda_2}{\lambda_\alpha \lambda_\beta}}
\end{equation}
It is evident from this definition that 
\begin{equation}
  \xi_{\alpha\beta} (x,\bar{z}_{\alpha\beta}) =  \xi_{\beta\alpha} (-x,\bar{z}_{\alpha\beta}) \neq  \xi_{\alpha\beta} (-x,\bar{z}_{\alpha\beta}),
\end{equation}
since for absorption by two clouds of gas at mean redshift
$\bar{z}_{\alpha\beta}$, 
the expectation value of the correlation is different for the case of a
lower-redshift cloud absorbing in $\alpha$ and a higher redshift cloud
absorbing in $\beta$ or vice-versa. As a result, the correlation
function is not symmetric around zero and the cross-power spectrum has
both real and imaginary components:
\begin{equation}
\xi_{\alpha\beta}(x,{\bar r}) = \frac{1}{2\pi}
\int_{-\infty}^\infty \left[P_{\alpha\beta}(k,\bar{z})+i Q_{\alpha\beta}(k,\bar{z})\right] e^{-ikx}dk.
\label{eq-xi-def}
\end{equation}
There exists no apriori argument that the imaginary part of the
cross-power spectrum should be zero. Since a non-zero
$Q_{\alpha\beta}$ reflects a non-symmetric problem it must be studied
for each specific case separately. In the next subsection we elaborate
why a non-zero $Q_{\alpha\beta}$ is expected in \lya-\lyb\ correlations.

A complete statistical description of \lya\ and \lyb\ fluctuations at
the two-point level is thus given by four power spectra
$P_{\alpha\alpha}$, $P_{\beta\beta}$, $P_{\alpha\beta}$,
$Q_{\alpha\beta}$. Each of these is a function of scale and
redshift.  

\subsection{Theoretical expectation for $P_{\beta\beta}$,
  $P_{\alpha\beta}$ and $Q_{\alpha\beta}$}
\label{sec:TeP}

Before proceeding, let us briefly discuss the expected quantities to be
measured by the new power spectra. 

First, the reader might be confused as to whether the new quantities are
truly linearly independent, since in the introduction we have argued that the
ratio of the optical depths is deterministic and known from atomic
physics. Indeed, they are independent for the following
reasons. Fluctuations in the optical depth are related to the
fluctuations in the transmitted flux fraction via a non-linear
transformation
\begin{equation}
\bar{F}(1+\delta_F) = e^{-\bar{\tau}(1+\delta_\tau) } =
e^{-\bar{\tau}}(1-{\bar \tau}\delta_\tau+\frac{1}{2}{\bar \tau}^2\delta_\tau^2 \ldots).
\end{equation}
We immediately see that $\bar{F}\ne e^{-\bar{\tau}}$, since even
zero-lag correlators contribute to the mean. Therefore, while
$\bar{\tau}_\beta=r_{\beta\alpha}\bar{\tau}_\alpha$, it is not
possible to write a similar relation between $\bar{F}_\alpha$ and
$\bar{F}_\beta$. By the same token, any 2-point statistics in
$\delta_F$ will contain contributions not just from the 2-point
statistics of $\delta_\tau$, but also all higher-order correlators
and hence one cannot write relations between $P_{\alpha\alpha}$ and,
for example, $P_{\beta\beta}$.

We do know, however, that on very large scales in
three-dimensions, both absorptions become linear tracers of the
underlying density field. Consequently one expects the 3D
cross-correlation coefficient to be close to unity 
\begin{equation}
r_{\rm 3D} (k)=\frac{P_{3D\ \alpha\beta}(k)}{\sqrt{P_{3D\ \alpha\alpha}(k)
    P_{3D\ \beta\beta}(k)}}\sim 1 \mbox{\ for\ small }k.
\end{equation}
Of course, due to stochasticity in the biasing relation (taking form of 
white noise in the low $k$ limit), the
cross-correlation coefficient will be somewhat less than unity,
but this effect is expected to be small (the absorption is, after all, coming
from exactly the same structure along each line of sight).

More importantly, however, the 1D power spectrum aliases small-scale
three-dimensional modes into large scale one-dimensional modes
\begin{equation}
P_{1D}(k) = \frac{1}{2\pi}\int_k^\infty P_{3D}(k') k' dk'. 
\end{equation}
Therefore the cross-correlation coefficient between \lya\ and
\lyb\ 1D power spectra, defined as
\begin{equation}
  \label{eq:rr}
  r =  \left[ \frac{P_{\alpha\beta}^2(k)+Q^2_{\alpha\beta}(k) }{P_{\alpha\alpha}(k)P_{\beta\beta}(k)}\right]^{1/2},
\end{equation}
is expected to be somewhat smaller than
unity, but one would not expect $r\ll 1$ at small $k$. 

Finally, in a non-evolving universe, $Q_{\alpha\beta}=0$.
The real Universe is evolving, but sufficiently slowly so that for small
separations the approximation of stationary statistics is in general
accurate. Hence, we expect $Q_{\alpha\beta}$ to be smaller than
$P_{\alpha\beta}$, i.e., the cross-power spectrum to be approximately
real. However, $Q_{\alpha\beta}$ is required for a statistically
consistent complete 
description of fluctuations in a given spectrum and thus it should be
measured together with other quantities.

\begin{figure}
  \centering
  \includegraphics[width=\linewidth]{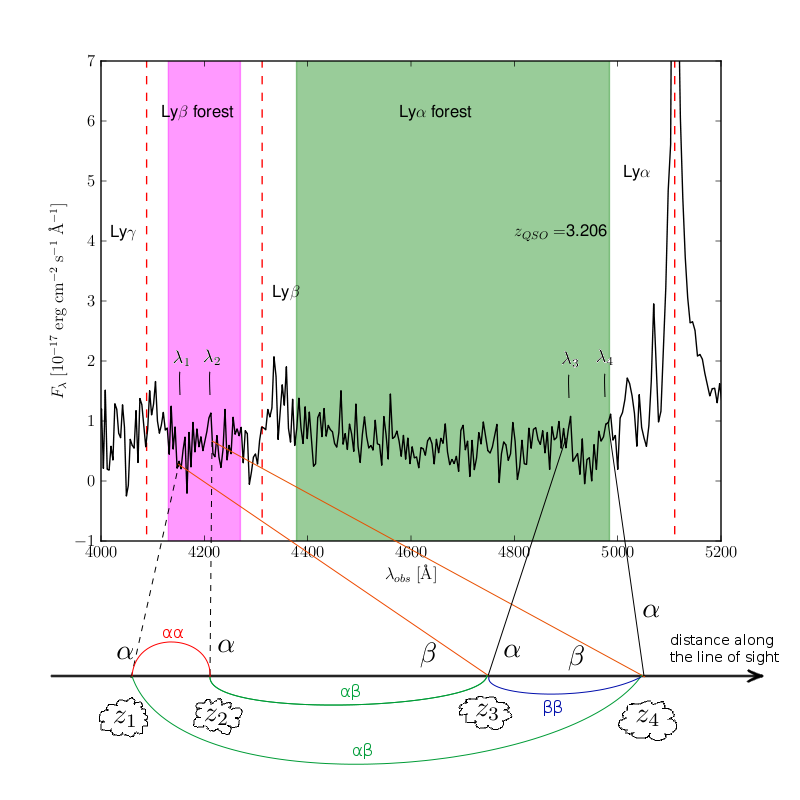}
  \caption{Geometry of the absorption in the $\alpha$ and $\beta$
    forests. A ``cloud'' of gas absorbing at redshift $z_3$ in \lya\ is
  also absorbing in the \lyb\ forest. However, the
  same pixel in the \lyb\ forest is also subject to absorption by
  another ``cloud'' at redshift $z_1$. Ditto for clouds at $z_4$ and
  $z_2$. When cross-correlating two pixels residing in the $\beta$
  region of the quasar spectrum, one must take into account four
  contributions to the correlations. When cross-correlating a pixel in
  the \lyb\ forest with one in the \lya\ forest, one must take into
  account two correlations.}
  \label{fig:niceplot}
\end{figure}

\begin{figure}
  \centering
  \includegraphics[width=1.0\linewidth]{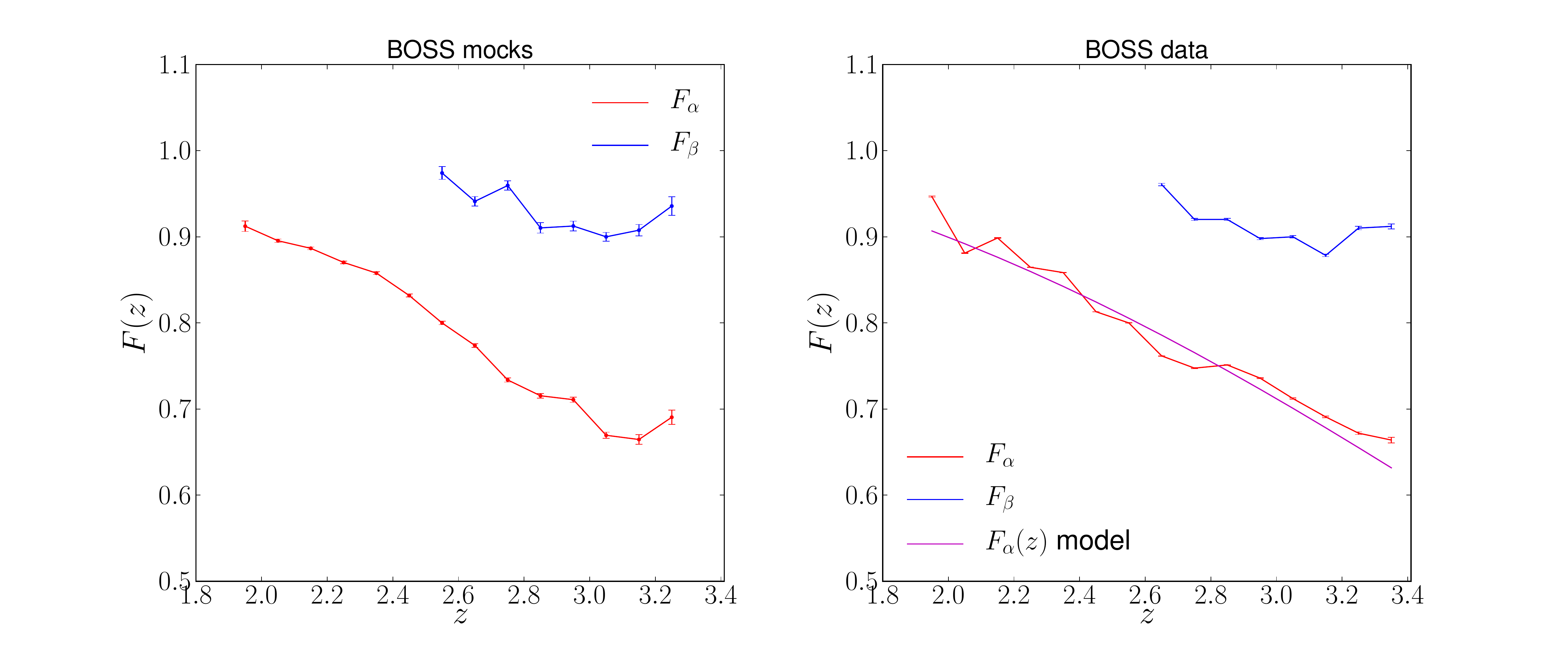}
  \caption{Mean flux of the \lya\ and \lyb\ fields inferred from the
    mock data set (left) and from the real data (right). The error
    bars on both plots are underestimated since they assume
    independent forest pixels. We overlay a curve
    $1.05\exp\left[-0.0046\left(1+z\right)^{3.3}\right]$ which
    describes data by \cite{2013arXiv1306.5896P} very well. The
    pre-factor of $1.05$ absorbs a different normalization between
    mean continuum and mean flux absorption, which are completely
    degenerate.  }
  \label{Fig:m1}
\end{figure}

\begin{figure}
  \centering
  \includegraphics[width=1.0\linewidth]{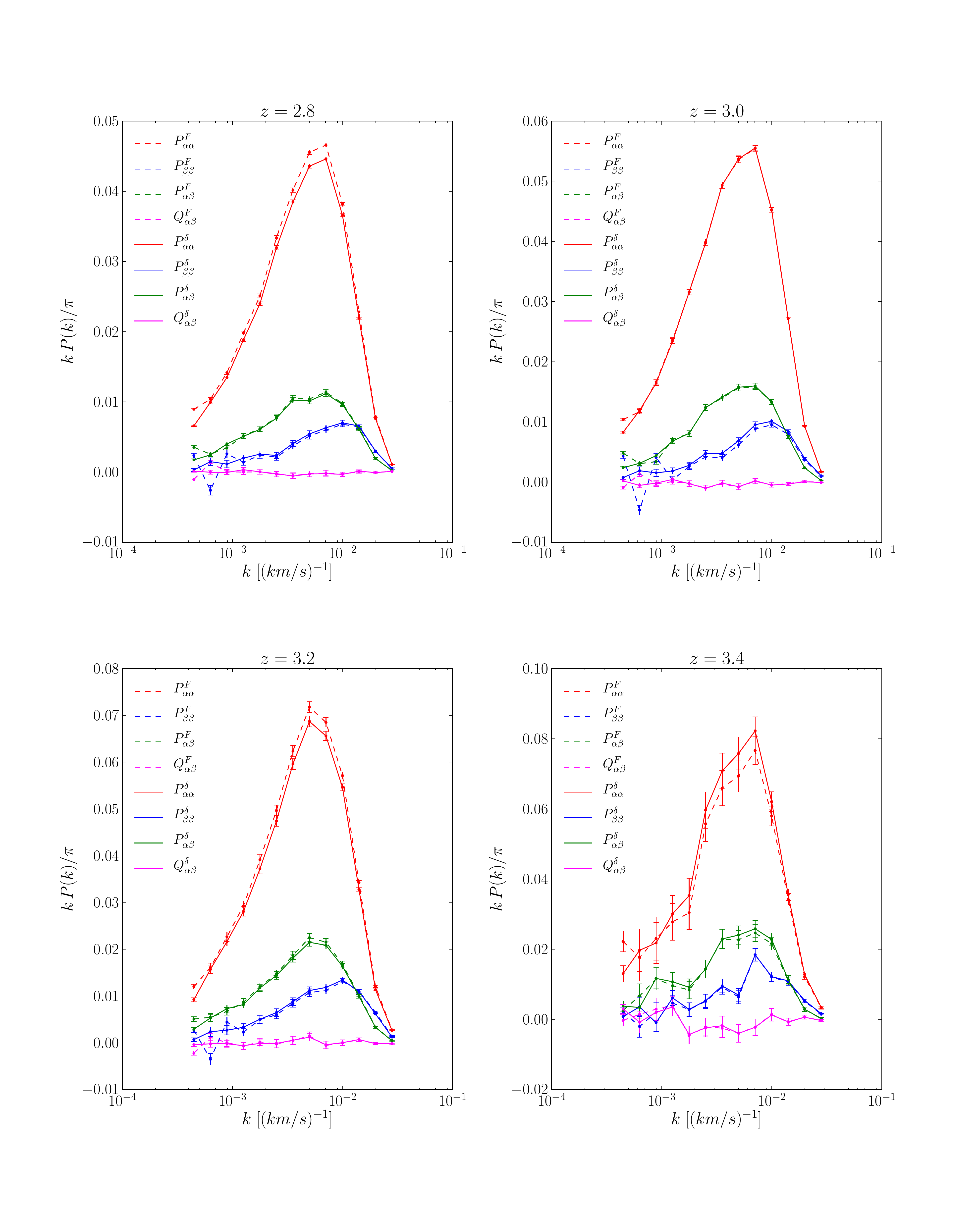}
  \caption{Power spectrum components measured on two different mock
    data sets: results with known quasar continua (solid line) and the
    full analysis (dashed line). This plot displays the mean of 10
    realizations of 10000 QSO mock data set. No PSF deconvolution has
    been performed and hence the power drops to zero at large $k$ values.}
  \label{Fig:m2}
\end{figure}

\begin{figure}
  \centering
  \subfigure[]{
    \includegraphics[width=0.47\linewidth]{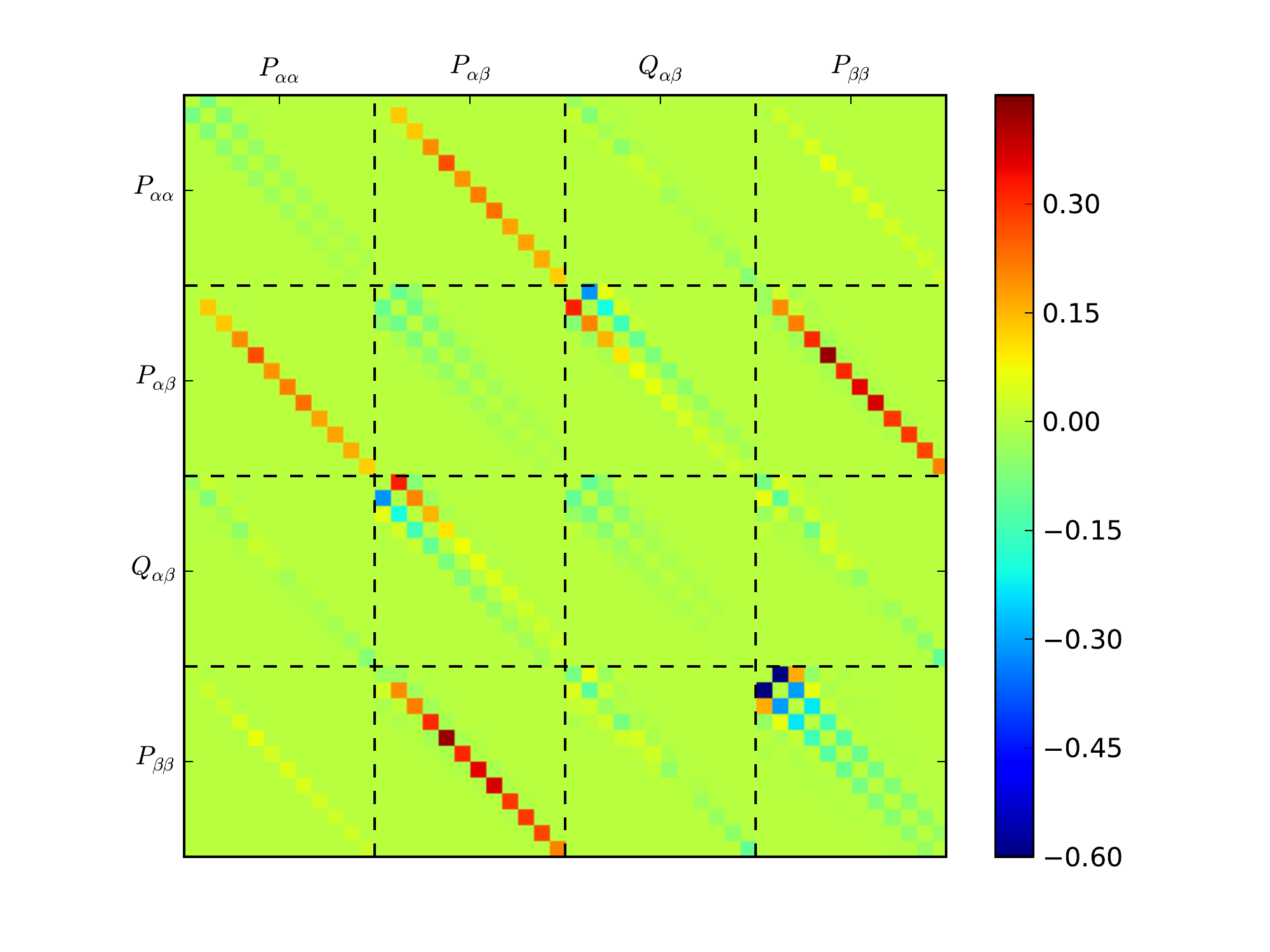}
    \label{subfig:f1}
  }
  \subfigure[]{
    \includegraphics[width=0.47\linewidth]{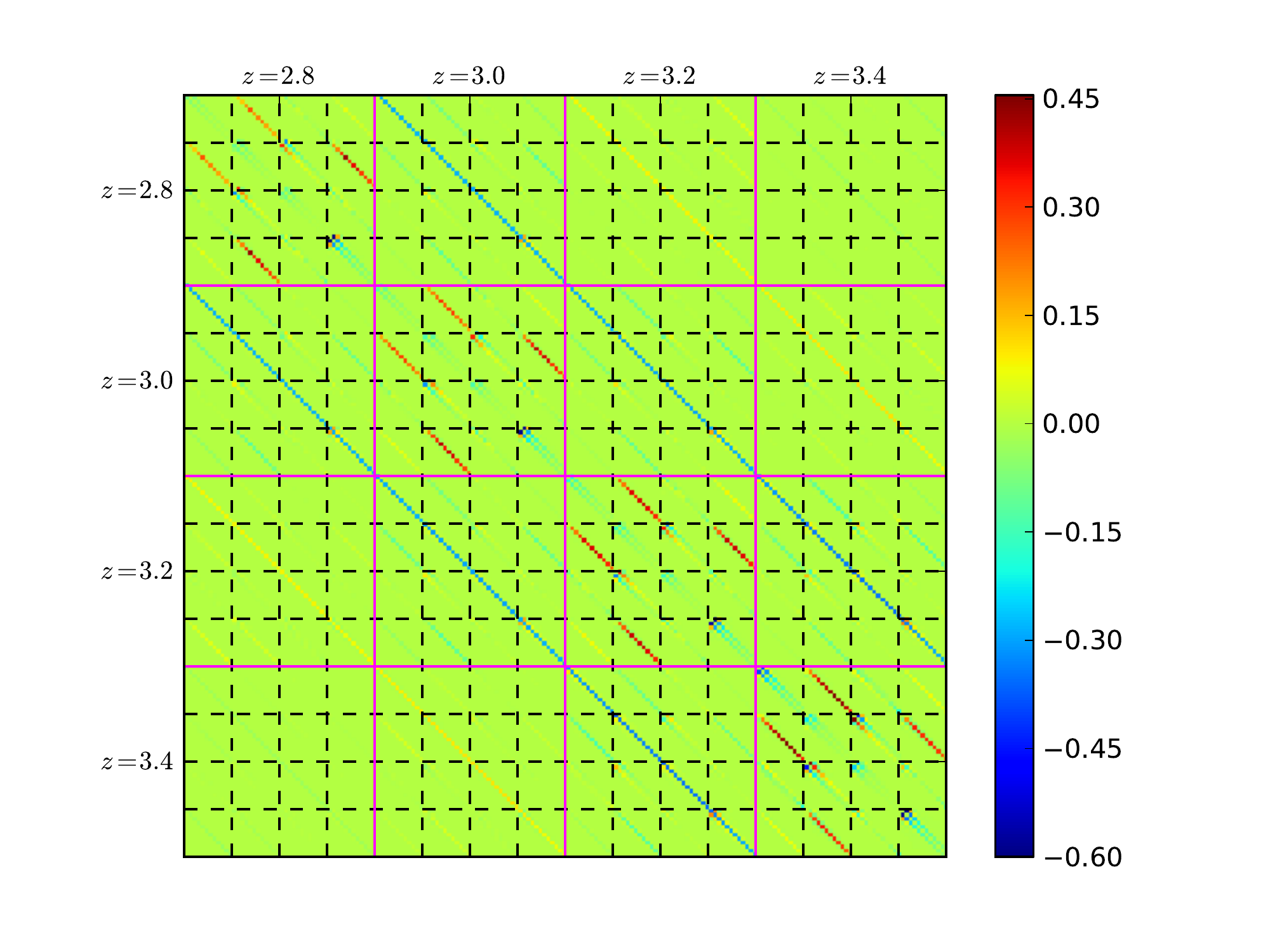}
    \label{subfig:f2}
  }

  \subfigure[]{
    \includegraphics[width=0.47\linewidth]{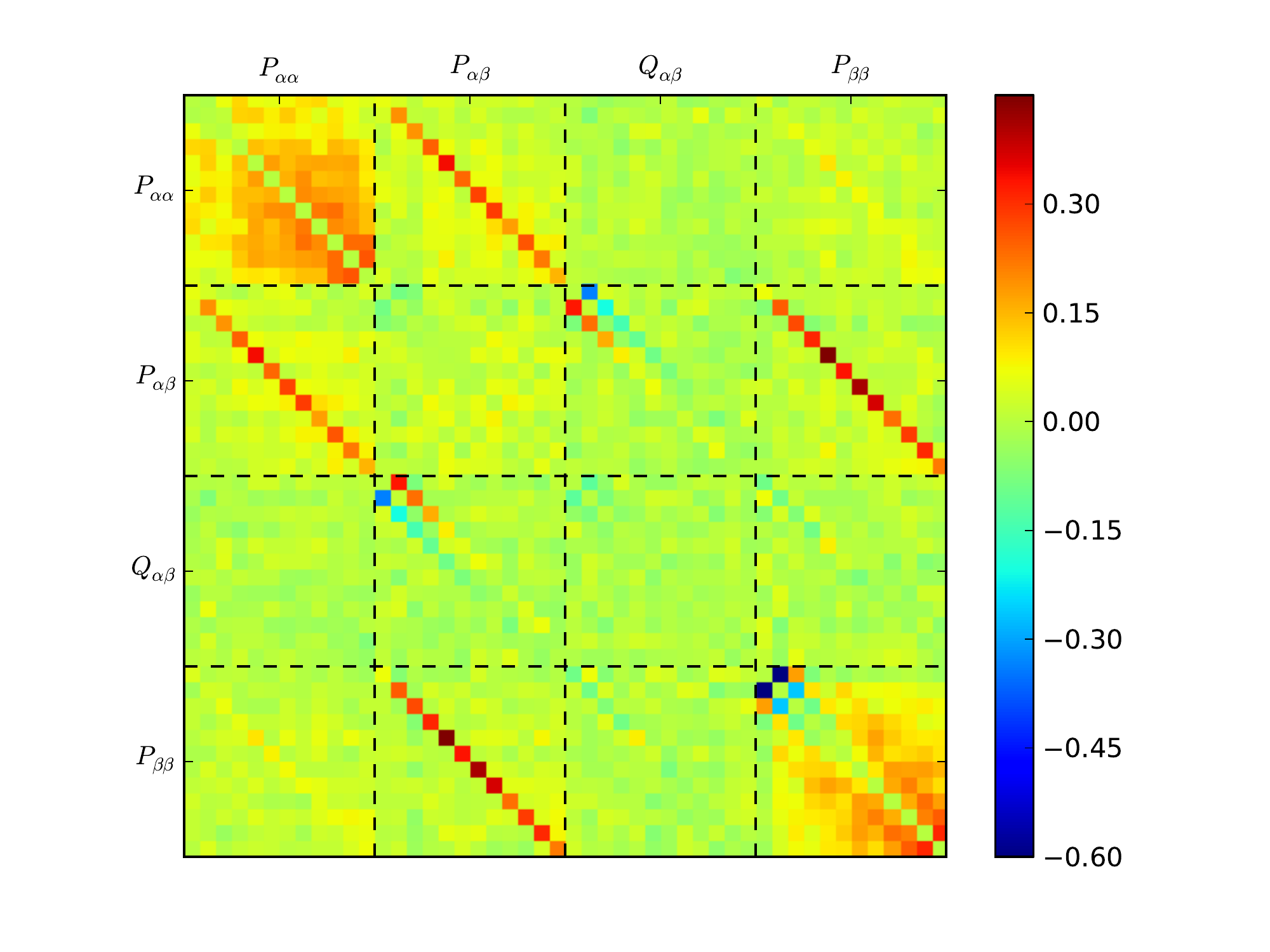}
    \label{subfig:f1b}
  }
  \subfigure[]{
    \includegraphics[width=0.47\linewidth]{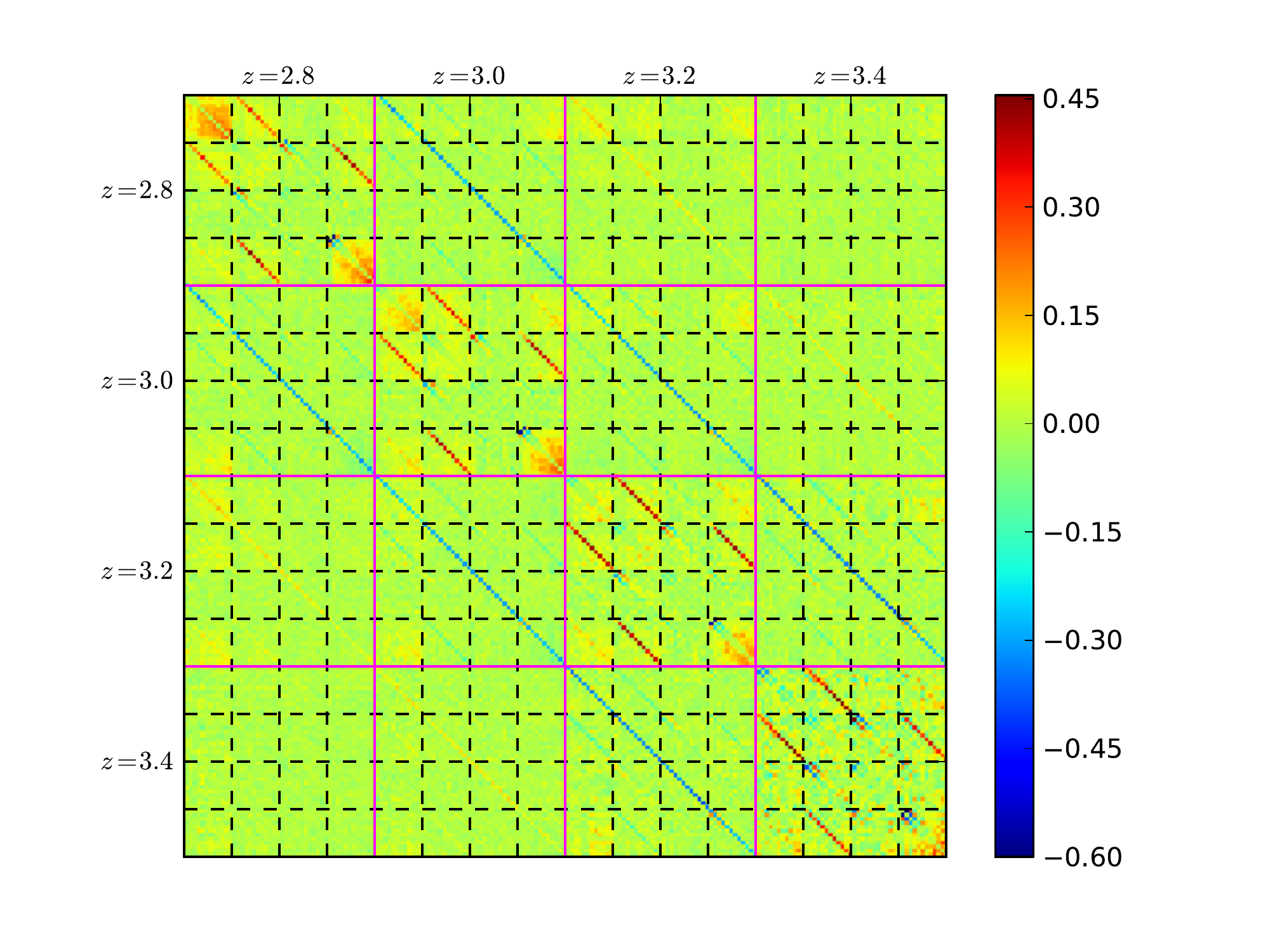}
    \label{subfig:f2b}
  }

  \caption{Error correlation matrices
    $\mathcal{C}_{ij}=C_{ij}/\sqrt{C_{ii}C_{jj}}$.  Top row figures (a
    and b) are for estimator covariance matrix, while bottom row are
    for the bootstrap derived covariance matrix. The left side figures
    (a and c) are an expanded view of the sub-matrix at redshift
    $z=2.8$ (upper left corner of the full matrix),
    while the right figures are the full matrices. The diagonal
    elements (unity) by definition were set to zero to increase the
    contrast. See text for discussion.}

  \label{Fig:cov}
\end{figure}

\subsection{Measuring power spectra from the data}
\label{sec:meas-power-spectra}
We proceed by discussing the reconstruction of these power spectra from
the data. The $P_{\alpha\alpha}$ can be extracted relatively directly,
but other components are more difficult, because $\beta$ absorption
is always contaminated by the lower redshift $\alpha$ absorption.

The model that we use for the observed quasar spectrum is given
by:
\begin{equation}
f^q(\lambda_i) = A^q {\bar C}(\lambda_i^{rest}){\bar F}_T(z_i)\left( 1 +
\delta_T(\lambda_i)\right).
\end{equation}

The continuum in each quasar $C^q(\lambda_r)$ is modeled by a
quasar amplitude $A^q$ and the mean continuum
$\bar{C}(\lambda_i^{rest})$. The absorption field is decomposed into a
mean absorption
\begin{equation}
  \bar{F}_T=
\begin{cases}
\lambda_r>\lambda_\alpha & 1 \\
\lambda_\alpha >\lambda_r>\lambda_\beta & F_\alpha(z) \\
\lambda_\beta >\lambda_r & F_\alpha(z) F_\beta(z) \\
\end{cases}
\end{equation}
and fluctuations
\begin{equation}
 1+\delta_T=
\begin{cases}
\lambda_r>\lambda_\alpha & 1 \\
\lambda_\alpha >\lambda_r>\lambda_\beta & 1+\delta_\alpha \\
\lambda_\beta >\lambda_r &  (1+\delta_\alpha)(1+\delta_\beta). \\
\end{cases}
\end{equation}
In this work, we ignore the second order contributions in the
\lyb\ forest 
\begin{equation}
  \delta_T(\lambda_o)=
  \delta_\alpha(r_\alpha)+\delta_\beta(r_\beta) +
  \delta_\alpha(r_\alpha)\delta_\beta(r_\beta) 
   = \delta_\alpha(r_\alpha)+\delta'_\beta(r_\beta)
\label{eq:2nd-order}
\end{equation}
and thus  work with effective fluctuations in the $\beta$ forest
\begin{equation}
  \delta'_\beta(r_\beta) = \delta_\beta(r_\beta) + \delta_\beta(r_\beta) \delta_\alpha(r_\alpha).
\end{equation}
Note that while quadratic term cannot be neglected, because it is not
small, it is for all practical purposes uncorrelated with the $\beta$
forest as it corresponds to gas that is $\sim 400 \,\mathrm{Mpc}$/h away -- on
scales considerably larger than the largest scales on which we measure
the power spectrum. Under this approximation the second term in the
above equation averages to zero. The cross correlation
$\langle \delta'_\beta \delta_\alpha \rangle \sim \langle \delta_\beta
  \delta_\alpha\rangle=\xi_{\alpha\beta}$, but the auto-correlation
gains an additional ``noise'' term $\langle \delta'_\beta
  \delta'_\beta\rangle \sim \xi_{\beta\beta} +
\xi_{\beta\beta}\xi_{\alpha\alpha}$. The $\alpha$ auto-correlation is
evaluated at the same distance separation $r=\Delta \log \lambda$, but at a lower redshift
$z_{\alpha}$
\begin{equation}
  1 + z_{\alpha} = \frac{\lambda_\beta}{\lambda_\alpha}\left(1 + z \right).
\end{equation}
Using the definition of the Fourier transform between correlation
function the power spectrum from Equation (\ref{eq:2}) the corrected
power spectrum can be written as
\begin{equation}
P_{\beta\beta}'(k,z) = P_{\beta\beta}(k,z) + \frac{1}{2\pi}
P_{\beta\beta}(k,z) * P_{\alpha\alpha}(k,z_{\alpha}) =
P_{\beta\beta}(k,z) + \frac{1}{2\pi}\int_{-\infty}^{\infty}
P_{\beta\beta}(y,z) P_{\alpha\alpha}(k - y,z_{\alpha}) dy,
\end{equation}
where ($*$) stands for convolution. Assuming $P_{\alpha\alpha}$ and
$P_{\beta\beta}$ to be approximately white, i.e.,
$P_{\alpha\alpha}(k,z) = \sigma_{\alpha}^2(z)$, where $\sigma^2$ is
the variance of the field, the correction to the cross correlation
coefficient is
\begin{equation}
r_{\alpha\beta}'(k,z) = r_{\alpha\beta}(k,z) \frac{1}{\sqrt{1 +
    \sigma^2_{\alpha}(z_{\alpha})}} \approx r_{\alpha\beta}(k,z)
\left( 1 - \frac{\sigma^2_{\alpha}(z_\alpha)}{2}\right).
\end{equation}

Since $z_\alpha$ is always smaller than $z$, for our highest measured
redshift bin of $z=3.4$, the corresponding $z_\alpha$ would be
$z_\alpha = 2.74$. The variance at $z_\alpha=2.74$ is approximately
$\sigma^2_{\alpha}(2.74) = 0.08$
(\cite{2013JCAP...04..026S,2011JCAP...09..001S}) and since the
variance is increasing with redshift
(\cite{2013JCAP...04..026S,2011JCAP...09..001S}) this is the largest
correction we would be able to apply. Thus, since the correction to
the cross-correlation coefficient is less than $5$\%, the effect is
well below what we can currently measure.  It is important to note,
however, that this will need to be carefully modeled for the future
precision observations.

Under these approximations, we now drop a prime on $\delta_\beta$ and
proceed with writing correlations between measured pixels in the
spectrum. A correlation of one pixel in the \lyb\ forest with a pixel
in the \lya\ forest is given by
\begin{equation}
\langle \delta_\alpha(\lambda_1) \delta_T(\lambda_2)  \rangle = 
\xi_{\alpha\alpha} (r^\alpha_2-r^\alpha_1) +
\xi_{\alpha\beta} (r^\beta_2-r^\alpha_1)
\end{equation}
and a correlation of two pixels in the \lyb\ forest contains four terms
\begin{equation}
\langle \delta_T(\lambda_1) \delta_T(\lambda_2) \rangle = 
\xi_{\alpha\alpha} (r^\alpha_2-r^\alpha_1) +
\xi_{\beta\beta} (r^\beta_2-r^\beta_1) +
\xi_{\alpha\beta} (r^\beta_2-r^\alpha_1) +
\xi_{\alpha\beta} (- (r^\beta_1 - r^\alpha_2)).
\label{eq:2ptxi}
\end{equation}

This is illustrated schematically in the Figure
\ref{fig:niceplot}. 

In this study we work within an optimal quadratic estimator framework
using the same methodology (and code-base) as in
\cite{2013JCAP...04..026S}. In particular, we model the power
spectrum functions $P$ and $Q$ as flat power-bands measured in $20$
bins, from $k = 0.000445881$ to $k = 0.05$ in steps of $\log k =
0.1$.  The lowest $k$ bin was extended to $k=0$.  In
redshift-direction, we use uniformly-spaced redshift bins from $z=1.9$
to $z=3.5$ in steps of $0.2$ and the model interpolates between values
determined at those redshifts. For \lyb\ and the cross power spectrum we
use redshift bins from $z=2.5$ to $z=3.5$. The redshift
corresponds to the true gas redshift, so \lyb\ absorption from a clump
of gas at $z<2.5$ is shifted into UV and thus not recorded by the
SDSS-III data. There is little signal in
lowest and highest redshift bins, but due to interpolation, we can
recover some information. 

In this parametrisation, the \lyb\ forest receives linear contributions
from all power spectrum bins.  Even in the case of
the usual \lya\ forest alone, however, a pair of pixels receives contributions
from all power spectrum bins, and hence from the point of view of a
quadratic estimator, our situation is not very different from the
standard case.

In short, the basic data-analysis proceeds as follows:
\begin{itemize}
\item We start by measuring the mean continuum and absorption as
  described in \cite{2013JCAP...04..026S}. This process has been
  extended to allow for an additional mean absorption in the $\beta$
  forest $\bar{F}_\beta$, but is otherwise the same as \cite{2013JCAP...04..026S}.
\item We then measure only \lya\ forest power spectrum using the
  mean continuum 
  and absorption from above. This provides a good starting estimate when
  measuring all the power spectrum components.
\item Lastly we measure all four power spectra $P_{\alpha\alpha},
  P_{\beta\beta}, P_{\alpha\beta}, Q_{\alpha\beta}$. This procedure is
  similar to the one described in \cite{2013JCAP...04..026S} but
  extended to \lyb\ region.

\end{itemize}

\subsection{Metal contamination at small velocity separations}
\label{sec:metal-cont-at}


In \cite{2005ApJ...635..761M}, it was found that absorption by \Si
contaminates the flux power spectrum measurement. \Si absorbs at a
wavelength $1206.50$\AA, which is close to the \lya\ absorption
wavelength, therefore \Si ``shadows'' the \lya\ correlations in the
forest. In principle, one could treat the \Si absorptions in
exactly the same manner as the \lyb\ absorptions - by writing a full
model for this contamination. 

While this is possible, it is certainly not easy, because any
estimator will have a difficult time distinguishing between the two
absorptions. The most likely result would be heavily correlated
measurements between \Si and \lya\ power. Therefore, it is easier to
treat the \Si absorption as a small correction to the \lya\ absorption.

We will later find a similar contamination issue in the \lyb\ forest. Both
contaminations leak power into $P_{\alpha\beta}$ and
$Q_{\alpha\beta}$. Fortunately, the cross-correlations are able to
distinguish between the relative signs of these absorptions.

We therefore develop a simple model with one contaminant in
the \lya\ forest and one dominant in the \lyb\ forest.

The basic assumption of this model is that the fluctuations of the
metal contaminant can be modeled as a scaled and shifted flux
fluctuation field of the \lya\ (or \lyb) field \cite{2005ApJ...635..761M}
\begin{equation}
\delta_\alpha'(x) = \delta_\alpha(x) + \delta_M(x) = \delta_\alpha(x) + a\delta_\alpha(x + v_\alpha).
\end{equation}
As discussed in the Section \ref{sec:TeP}, this approximation does
eventually break down at some level of precision, but it does provide
a good fit to the data. For a more detailed analysis of metal
contaminations see \cite{2012JCAP...07..028F}.

This model of flux fluctuations yields the following power spectrum
\begin{equation}
P_{\alpha\alpha}'(k) = P_{\alpha\alpha}(k) \left[ 1 + a^2 + 2 a \cos\left(k v_\alpha\right) \right].
\end{equation}
and ditto for the \lyb\ power spectrum contaminated with a metal of
strength $b$ and frequency $v_\beta$.

In the cross-power spectrum, this model affects both the real and imaginary
components of the cross power spectrum
\begin{align}
P_{\alpha\beta}'(k) &= n(k) P_{\alpha\beta}(k) + m(k) Q_{\alpha\beta}(k), \\
Q_{\alpha\beta}'(k) &= n(k) Q_{\alpha\beta}(k) - m(k)
P_{\alpha\beta}(k),
\label{eq:Qcrt}
\end{align}
where the functions $n(k)$ and $m(k)$ are given by
\begin{align}
n(k) &= 1 + a\cos\left(k v_\alpha\right) + b\cos\left(k v_\beta\right)
+ ab\cos\left[k \left(v_\beta - v_\alpha\right)\right], \\
m(k) &= a\sin\left(k v_\alpha\right) - b\sin\left(k v_\beta\right)
- ab\sin\left[k \left(v_\beta - v_\alpha\right)\right].
\end{align}
The metal contaminant mixes the intrinsic real and
imaginary part of the cross power. This means that even if there would
have been no intrinsic imaginary power one would still measure
non-zero contribution of the imaginary cross power spectrum. This
conclusion makes sense intuitively. $Q_{\alpha\beta}=0$ requires the
distribution of \lya\ and \lyb\ absorptions to be symmetrical with
respect to the inversion of the radial axis; a metal absorption at a
small separation with a fixed sign will naturally break this symmetry.

Finally, we note that \emph{for this particular model} of metal
contamination, the contamination cancels perfectly for the
cross-correlation coefficient defined as in Equation \ref{eq:rr}.

\section{Data \& Synthetic data}
\label{sec:data}

In this work we use BOSS quasars from the Data Release 9 (DR9;
\cite{2012ApJS..203...21A}) sample. The quasar target selection for
the DR9 sample of BOSS observations is described in detail in
\cite{2012ApJS..199....3R} and we refer reader to that publication for
the details. 

We model continuum over the rest frame wavelength range of $978$\AA\
to $1600$\AA. This region is the same as in \lya\ analysis of the paper Slosar
et al. (\cite{2013JCAP...04..026S}) but is extended to lower rest frame
wavelengths to enclose the \lyb\ forest. For the purpose of our
analysis we define the \lya\ forest to be $1041 - 1185$\AA, which
is similar to the range used by McDonald et
al. (\cite{2006ApJS..163...80M}) and more conservative than the range
in \cite{2013JCAP...04..026S}.  The upper limit for the \lya\ forest is
thus roughly in the regime where proximity effects and \lya\ emission
line profile can be assumed to be small. For similar reasons, the
lower limit is also  kept a safe distance away from the \lyb\ emission
peak.

In similar spirit we define the \lyb\ forest region as rest frame $978
- 1014$\AA. This range is a bit more
conservative than the \lya\ range, since the Lyman emission peaks
become narrower as one moves long the series (i.e. \lyb\ emission peak
is narrower than \lya\ emission peak). The \lyb\ forest
range covers a much shorter path length than the \lya\ region, which
means inherently less signal. Also, we reiterate that while there is
only \lya\ absorption in the \lya\ forest region defined in this
paper, there are both \lya\ and \lyb\ absorptions in the \lyb\ forest
region. Of course, there is metal contamination throughout both forests.

\subsection{Mock data}

We tested our technique on the same mock data as used in
\cite{2013JCAP...04..026S, 2012JCAP...01..001F}. It is important to
stress this mock data-set is not optimal for testing this analysis, since it
is focused on the three-dimensional correlations. The small scale
power is roughly correct, but only at an order-of-magnitude level.
Since we are not aiming at precision cosmology, this should not be a
major handicap for our study. If one demonstrates that we can measure the power
spectrum without a major bias in these mock data-sets, we are also
likely to be making reliable measurements in the true data.

To extend the mock data used in \cite{2013JCAP...04..026S} to the
\lyb\ forest, we scaled the optical depth in the $\tau_\alpha$ field by
$r_{\beta\alpha}$ (see Equation (\ref{eq:3})) and translated it to an
appropriate redshift. The Principal Component Analysis (PCA) continua
do not extend to these low redshifts and so we artificially extend
them with a constant value.

\section{Application to mock data}
\label{sec:mock}

We tested our analysis on the mock data as follows. First, we
demonstrated that our quadratic estimator yields an unbiased
result for a white noise input signal which perfectly cross-correlates
$\alpha$ and $\beta$ fields.  Next we applied our estimator to the
mock data-set, assuming perfect knowledge of the continuum and mean
absorption. These results were compared with the full analysis, in
which we infer all the quantities from the data, as we must do
with the real data.  We present these results in Figures \ref{Fig:m1} and
\ref{Fig:m2}. 

Figure \ref{Fig:m1} shows the inferred mean absorption from the
mock data-set for both \lya\ and \lyb, together with real measurements
discussed in the next section. For this section, the relevant plot is
Figure \ref{Fig:m2}, which shows how fitting for the continuum
fluctuations affects the measured power spectra. Small disagreements
are consistent with  the fact that the mock data-set misreports the noise-levels
to mimic our real misunderstanding of the noise properties of
spectrograph (see section 2.2 of \cite{2013JCAP...04..026S}). We have
also performed simpler tests for which we assumed the \lya\ forest
field to be perfectly white, fit the data with a single power spectrum
bin and compared this with direct estimates using variances - this
test convinced us that we do not have missing pre-factors in our
estimator. However, we have not carefully tested
redshift-interpolation and other more subtle aspects of the estimator.

These tests lead us to conclude that our data analysis will be able
to reconstruct the measured power spectrum at the level of precision
relevant for this exploratory work when applied to the real data. 

\section{Results}
\label{sec:result}

We applied our data reduction method to the data.  Much of the
analysis is common with \cite{2013JCAP...04..026S} and we refer the reader
to that publication for more details. In Figure \ref{Fig:m1} we plot
$\bar{F}$ for \lya\ and \lyb\ forests in mock data and real data.
The absolute normalization of each individual mean absorption is
arbitrary (since it is degenerate with the mean continuum
shape in the relevant forest regions). The error bars are
underestimated, since they do not correctly take into account
correlations between pixels. Nevertheless, the visual agreement
between the results on the mock data and real data is quite good and, in
fact, better than one would naively expect given that the small scale
power is not appropriately reproduced in these mock data.

Next we discuss the covariance matrix of our measurements.  In the top
row of Figure \ref{Fig:cov} we show the covariance matrices derived
from the optimal estimator and the data.  The covariance matrix has
the expected structure. Measurements of the \lya\ power spectrum are
effectively uncorrelated, with only weak anti-correlation between
adjacent bins. Measurements of the \lyb\ power spectra are similar,
but the anti-correlations between adjacent bins are larger, since the
available path length is smaller. Measurement of the cross-power
spectra are also only weakly internally correlated, but they show
significant correlations with both auto power spectra. The most
interesting aspect is the covariance structure of the $Q_{\alpha\beta}$ with
$P_{\alpha \beta}$, where bins at the same $k$ are uncorrelated, but
are somewhat correlated with adjacent $k$-bins.

Measurements of the 1D quantities in the data are conveniently
bootstrapped by assuming each quasar to be an independent measurement
of this quantity (this should be an excellent approximation).  We
generated 3000 bootstrap samples of our dataset and calculated the
corresponding bootstrap covariance matrix.  When compared with the
bootstrap derived covariance matrix, the estimator under-estimates
the diagonal elements of the covariance matrix by approximately
10\%. We correct for this error in subsequent use of the matrix by
multiplying all element of the covariance matrix by 1.1.  The
correlation structure for this matrix is displayed in the bottom row of Figure
\ref{Fig:cov}.  We see that compared to the estimator matrix, the
structure is in general similar. One important difference is that the
bootstrapping is selecting a constant-like contribution to variance in
the auto-correlations. This feature likely arises due to our imperfect
fitting of the quasar amplitude for small signal-to-noise quasars that
modulates the power spectrum normalization.

\begin{figure}[H]
  \centering
  \includegraphics[width=1.0\linewidth]{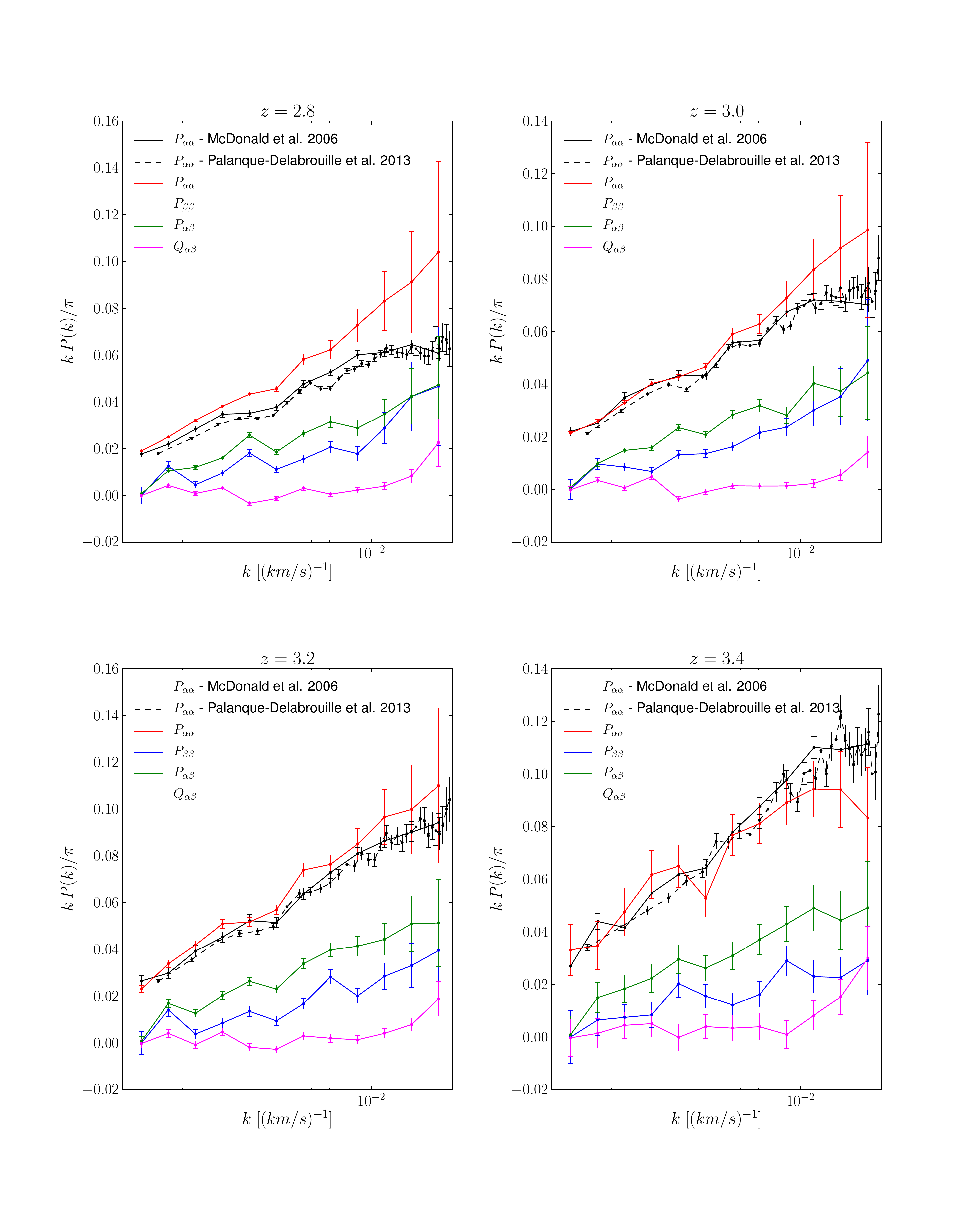}
  \caption{Measured power spectrum components: \lya\ power spectrum
    (red), \lyb\ power spectrum (blue), real (green) and imaginary
    (magenta) part of the cross power spectrum. It can be clearly seen
  that both \lya\ and real part of the cross power spectrum are
  detected with high significance while the imaginary part has lowest
  detection significance. Also apparent are oscillations in all four 
components. We compare our measurements with those by
\cite{2006ApJS..163...80M} and \cite{2013arXiv1306.5896P}. We have
added the background contribution to both of those measurements.
}
  \label{Fig:1}
\end{figure}

So far, for example in mock-testing, we have completely neglected the
effect of the finite spectrograph resolution and pixel size. Both
effects smooth the observed fluctuations and thus dampen the power on
small scales. To account for this effect properly, a correction has
to be used in the estimator that convolves the power spectrum and the
smoothing kernel for each $k$ bin of the power spectrum. Since for
this work we were interested only in a rough estimate we proceed to
make an approximate correction as follows. We estimate our beam
correction as a mean of the correction kernel over pairs of pixels
that contribute in the same $(k,z)$ bin. The beam correction we apply
is thus given by a weighted average
\begin{equation}
B(k) = \frac{\sum_{i,j \in \text{pairs}} w_i w_j W(\lambda_i)
  W(\lambda_j)}{\sum_{i,j \in \text{pairs}} w_i w_j},
\end{equation}
where the weights were given as inverse square variance for each
pixel. The smoothing kernel is given by (\cite{2006ApJS..163...80M})
\begin{equation}
W(k,\lambda_i) = \exp\left(-k^2 r_i^2\right) {\rm sinc}{\left(\frac{k p_i}{2}\right)},
\end{equation}
where $p_i$ is the pixel width for the pixel $i$ given by $\lambda_i$
and $r_i$ the resolution for the same pixel. The errors on the
spectrograph resolution $r_i$ are estimated to be of order of 10\%
(\cite{2013arXiv1306.5896P}). This results in a substantial increase
in the size of the error-bars in our measurements at high $k$.

In Figure \ref{Fig:1} we compare our measurements and the
measurements of the \lya\ forest alone using 3000 SDSS quasars by
\cite{2006ApJS..163...80M}. For comparison we also add the
measurements of the \lya\ forest only by a recent study of the new BOSS
release using 14000 BOSS quasars by \cite{2013arXiv1306.5896P}. 
In general, we find good agreement, except
at the lowest redshift bin, where we measure excess power when
compared to the other measurements. The most likely
explanation for mismatch is poor noise modeling in our data, since it is
known that the pipeline noise is not accurate \cite{2013JCAP...04..026S}.
 
\begin{figure}[ht]
  \centering
  \includegraphics[width=1.0\linewidth]{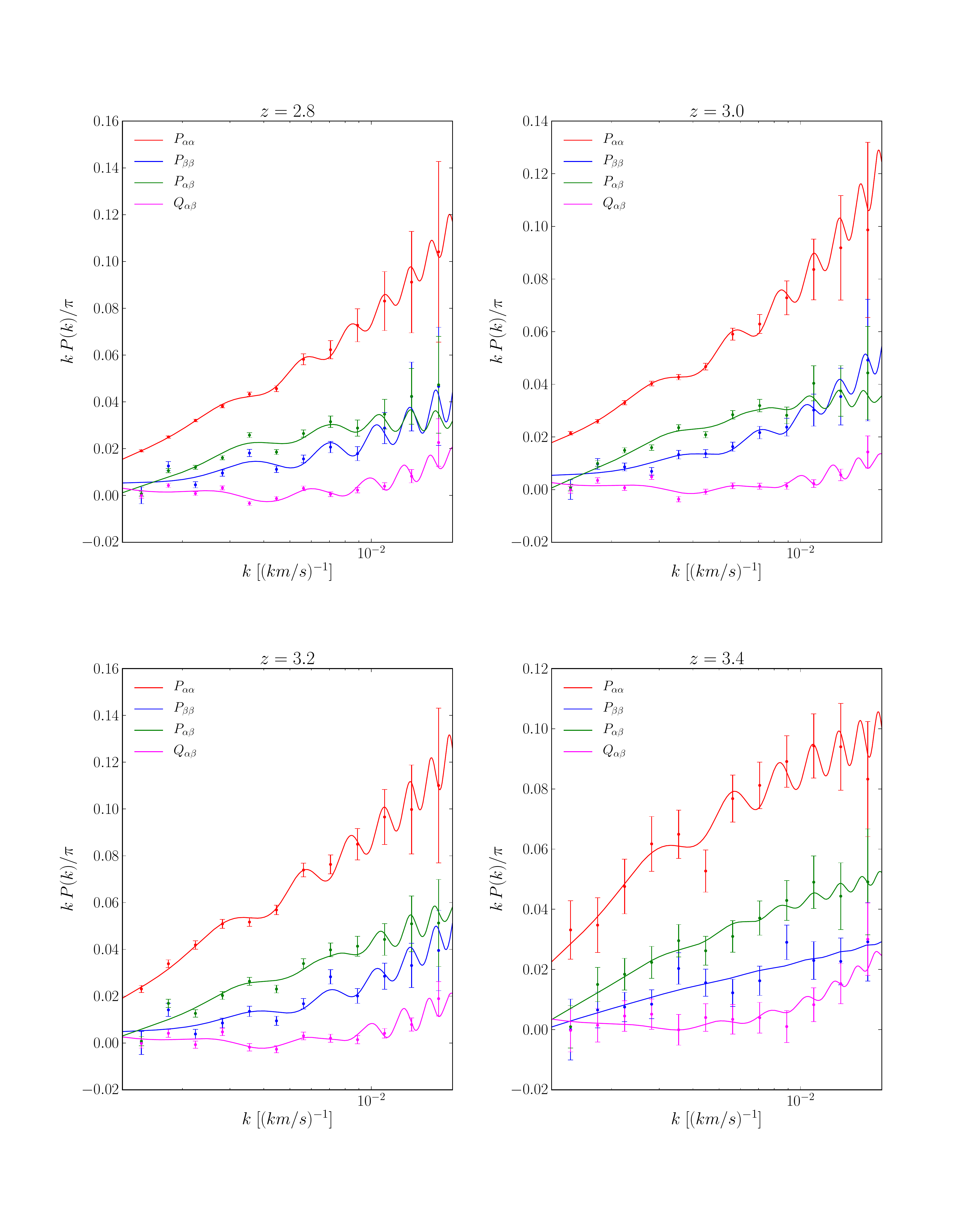}
  \caption{Fits for power spectrum models with metal
    contaminants in \lya\ and \lyb\ forest. The model does not produce a
  good fit to the data, but the oscillation frequencies are measured
  very robustly.}
  \label{Fig:12}
\end{figure}

\begin{figure}[ht]
  \centering
  \includegraphics[width=1.0\linewidth]{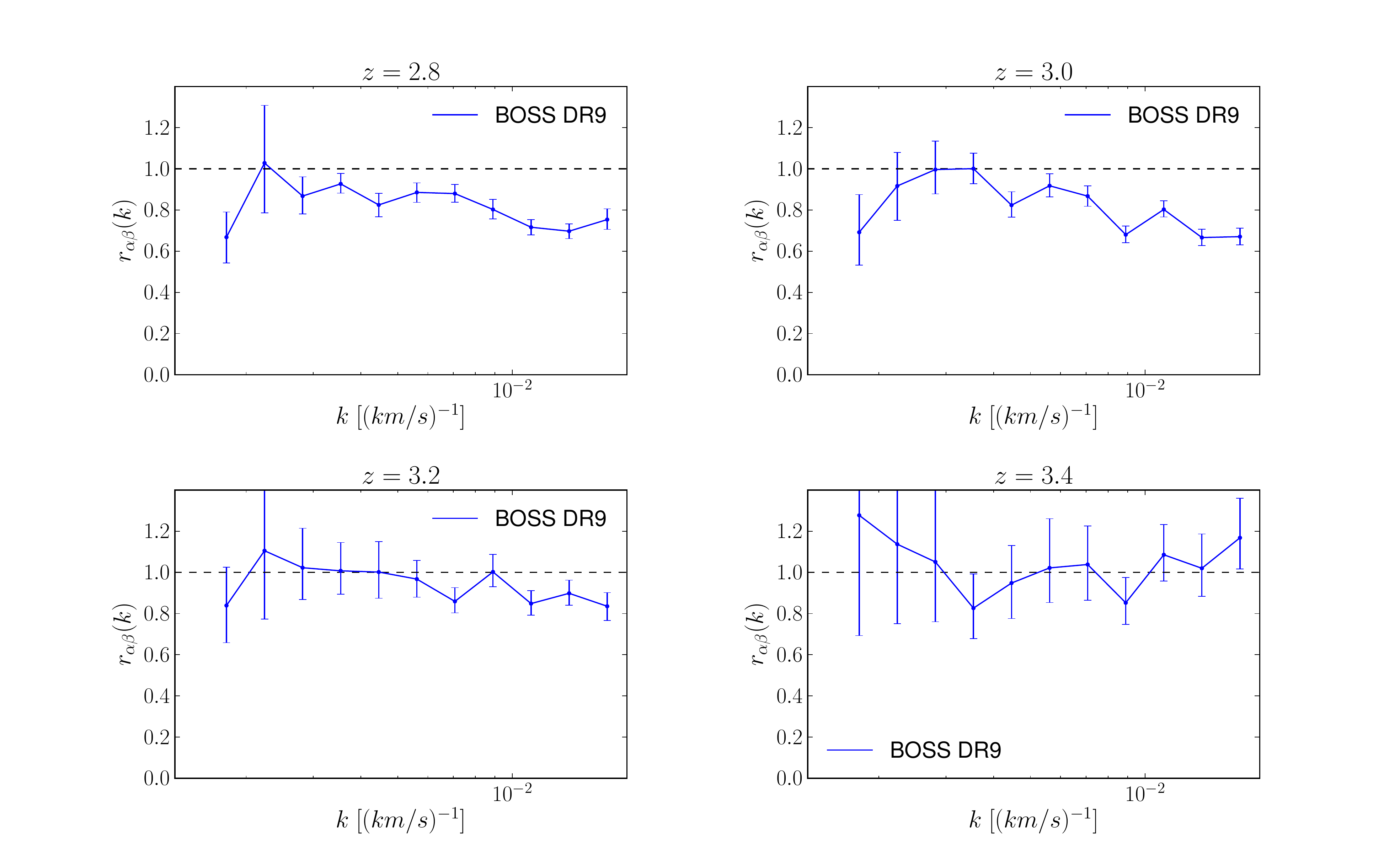}
  \caption{Measured cross correlation coefficient (\ref{eq:rr}) from the
    data. Within the error bars, the coefficient is constant on large
    scales and is falling off towards smaller scales.  }
  \label{Fig:2}
\end{figure}

Using the data plotted on Fig. \ref{Fig:1} we estimated the
significance with which we measure a non-zero imaginary part of the
cross power spectrum $Q_{\alpha\beta}$. For this estimation we have
only used modes with $k<0.01\;\mathrm{km\,s^{-1}}$. Significance
$s=\sqrt{\sum (Q_{\alpha\beta}/\sigma_{Q_{\alpha\beta}})^2}$ of
$Q_{\alpha\beta}$ in each redshift bin was estimated to be $(z,s)$:
$(2.8,8.24)$, $(3.0,7.23)$, $(3.2,4.97)$ and $(3.4,1.93)$. The total
significance of measuring $Q_{\alpha\beta}$ different from zero was
estimated to $12.2\sigma$. We caution reader that this significance
corresponds to the contaminated cross power spectrum component given
by Equation \ref{eq:Qcrt} and thus composes entangled information from
both the intrinsic imaginary part of the power spectrum and metal
contamination. 

From Fig. \ref{Fig:1} it is apparent that oscillations are
imprinted on top of a smooth power spectrum.  We propose that these
oscillations are best described as being due to the presence of
contaminating metal at small separation from the main absorption
line (\cite{2006ApJS..163...80M}).

In order to test this hypothesis, we fit our data as an intrinsically
smooth power spectrum described by a 2nd order polynomial fit in
$\log(k)$. All four components ($P_{\alpha\alpha}$, $P_{\beta\beta}$,
$P_{\alpha\beta}$, $Q_{\alpha\beta}$) were fit with independent
smooth component at each redshifts.  We convolved this model with a
dominant metal contamination at fixed separation as described in the
section \ref{sec:metal-cont-at}. We assumed that contaminating
oscillation strength is independent in each redshift bin, but that
the oscillation frequencies ($v_\alpha$ and $v_\beta$) are fixed.

To get an appropriate initial parameters for the optimizer, we first
used a simple model fitting only \lya\ power spectrum with one
contaminating metal. With this simpler model we explored a larger part
of the phase-space and determined a rough estimate for \lya\ frequency
to lay around $\pm 2000\;\mathrm{km\,s^{-1}}$ (note that auto power
spectra cannot determine the sign of the contaminating velocity). We
used the same simple model for \lyb\ power spectrum only and again
after exploring a large part of the phase-space for $v_\beta$ found a
rough estimate of around $\pm 1800\;\mathrm{km\,s^{-1}}$. We then
proceeded to use those as starting points, with four possible sign
permutations for a finer fitting with all available data, including
cross-correlations. Only the presented sign combination converged.

The best-fit model resulting from this procedure can be found
in Figure \ref{Fig:12}. This model did not produce a good fit to
the data - in fact our best fit gives $\chi^2=299.92$ with $134$
degrees of freedom (even after correcting for the 10\%
error-covariance underestimate). Not surprisingly, we have found that
the two robustly measured quantities are the oscillation frequencies
of the contaminating components, which are given by

\begin{align}
v_\alpha &= 2269 \pm 19 \; \mathrm{km\,s^{-1}}, \\
v_\beta &= -1820 \pm 13 \; \mathrm{km\,s^{-1}}.
\end{align}

Since we do not produce a good fit to the data, the error bars are likely
underestimated. Nevertheless, we can identify the contaminants.  The
metal contaminant in \lya\ forest ($v_\alpha$) is the \Si line
transition, absorbing at $1206.5$\AA, which is separated from \lya\ by
$v_\alpha=2271\, \mathrm{km\,s^{-1}}$ confirming results by
\cite{2006ApJS..163...80M}.  The contaminant in the \lyb\ forest is
identified with \Ox that absorbs at $1031.9$\AA, corresponding to
$v_\beta=-1801\, \mathrm{km\,s^{-1}}$.

We proceed by examining the cross-correlation coefficient defined in
Equation \ref{eq:rr}. As mentioned in Section \ref{sec:metal-cont-at},
under the simplified model of metal contamination, its effect cancels
exactly in this quantity. Errors due to
absolute noise power that affect the auto power spectra but not the
cross power spectra will, in general, affect this quantity. The
quantity $r_{\alpha\beta}(k)$ is plotted in Figure (\ref{Fig:2}). The statistical
error bars on this plot were derived by drawing samples of power
spectra consistent with the measured data and the associated
covariance matrix and examining the resulting scatter in $r$.
Although the measurements are uncertain and error bars large, the
general behavior follows the expectations. On large scale we see
nearly unity cross-correlations that tends to decrease towards smaller
scales.

\section{Conclusion}
\label{sec:conc}

In this paper we studied the possibility of measuring the
\lyb\  forest in spectra of quasars.  The fact that the
underlying density field evolves with redshift breaks the symmetry
along the line of sight when measuring cross power spectrum which
results in a cross-correlation function that is not symmetric with
respect to changing the sign of the velocity difference. This yields an
intrinsic non-zero imaginary component to the cross power
spectrum. When considering the \lyb\ in addition to \lya\ forest, one
therefore measures three new components. Including higher Lyman
transitions will add new auto power spectra and in general two new
cross-power spectra for any combination of absorbing lines. However,
due to decreasing path-length of higher-order forests, it is not clear
whether it is useful to venture beyond the \lyb\  line.

Measurements of the \lyb\ power spectrum and the \lya-\lyb\ cross power
spectra offer an improved way of estimating cosmological parameters
over using the \lya\ power spectrum alone, since we expect that many of
the astrophysical nuisance parameters that are degenerate with the
cosmologically interesting parameters can be measured
semi-independently from the new quantities. This stems from the fact
that the two transitions map the same intergalactic medium, but are
sensitive to different density and temperature ranges. This presents
an opportunity to better constrain IGM parameters of the flux-density
transformation and thus break the degeneracies between IGM
parameters (especially parameters of the equation of state) and
cosmology parameters (e.g. scalar spectral index).

Measurements of the cross power spectra $P_{\alpha\beta}$ are
independent of the choice of noise model. With future theoretical
modeling, we should be able to predict the cross-correlation coefficient
accurately and therefore the cross-power spectra will provide a
convincing self-consistency check.

We have measured the quantities discussed above in the BOSS DR9
data. Our work is clearly not accurate at the level required for
precision cosmology fits. In particular, effects of noise,
spectrograph resolution and metal contamination (both in-forest like
\Ox, but also lower redshift metals that are uncorrelated with the
signal of interest).  Along with a better data analysis, the theory
also needs to be further investigated using numerical simulations of
the \lyb\ forest. These are trivial to generate from the 
\lya\  simulations by appropriate rescaling of the optical depth. 

Nevertheless, we have measured power in all quantities discussed above
with high significance. Our measurements confirm the standard picture
describing the \lya\ forest. The cross-correlation coefficient is close
to unity on large scales as expected from qualitative arguments. We
found oscillations in all the power spectra measured. Our fits
indicate that these features are best explained by a combination of the \Si
contamination of the \lya\ forest (known previously) and \Ox contamination
in the \lyb\ forest (new to this work).

\section*{Acknowledgments}

Funding for SDSS-III has been provided by the Alfred P. Sloan
Foundation, the Participating Institutions, the National Science
Foundation, and the U.S. Department of Energy Office of Science. The
SDSS-III web site is \texttt{http://www.sdss3.org/}.

SDSS-III is managed by the Astrophysical Research Consortium for the
Participating Institutions of the SDSS-III Collaboration including the
University of Arizona, the Brazilian Participation Group, Brookhaven
National Laboratory, University of Cambridge, Carnegie Mellon
University, University of Florida, the French Participation Group, the
German Participation Group, Harvard University, the Instituto de
Astrofisica de Canarias, the Michigan State/Notre Dame/JINA
Participation Group, Johns Hopkins University, Lawrence Berkeley
National Laboratory, Max Planck Institute for Astrophysics, Max Planck
Institute for Extraterrestrial Physics, New Mexico State University,
New York University, Ohio State University, Pennsylvania State
University, University of Portsmouth, Princeton University, the
Spanish Participation Group, University of Tokyo, University of Utah,
Vanderbilt University, University of Virginia, University of
Washington, and Yale University.

\bibliographystyle{plain}
\bibliography{repo}


\end{document}

%% file: mydefinitions.tex
\newcommand{\be}{\begin{equation}}
\newcommand{\ee}{\end{equation}}

\newcommand{\lya}{Ly$\alpha$\ }


%% file: authors_fix.tex
\author[a]{Vid Ir\v{s}i\v{c},}
\author[b]{An\v{z}e Slosar,}
\author[c]{Stephen Bailey,}
\author[d]{Daniel J. Eisenstein,}
\author[e,c]{Andreu Font-Ribera,}
\author[f]{Jean-Marc Le Goff,}
\author[g]{Britt Lundgren,}
\author[c]{Patrick McDonald,}
\author[h]{Ross O'Connell,}
\author[f]{Nathalie Palanque-Delabrouille,}
\author[i]{Patrick Petitjean,}
\author[f]{Jim Rich,}
\author[f]{Graziano Rossi,}
\author[j,k]{Donald P. Schneider,}
\author[b]{Erin S. Sheldon,}
\author[f]{Christophe Y\`{e}che}

\affiliation[a]{Faculty of Mathematics and Physics, University of Ljubljana, Jadranska 19, 1000 Ljubljana, Slovenia}
\affiliation[b]{Brookhaven National Laboratory, Blgd 510, Upton NY 11375, USA}
\affiliation[c]{Lawrence Berkeley National Laboratory, 1 Cyclotron  Road, Berkeley, CA 94720, USA}
\affiliation[d]{Harvard-Smithsonian Center for Astrophysics, MS \#20, 60 Garden St., Cambridge, MA 02138, USA}
\affiliation[e]{Institute of Theoretical Physics, University of Zurich, Winterthurerstrasse 190, 8057 Zurich, Switzerland}
\affiliation[f]{CEA, Centre de Saclay, IRFU,  F-91191 Gif-sur-Yvette, France}
\affiliation[g]{Department of Astronomy, University of Wisconsin, 475 North Charter Street, Madison, WI 53706, USA}
\affiliation[h]{Physics Department, Carnegie Mellon University, Pittsburgh, PA 15213, USA}
\affiliation[i]{Universit\'e Paris 6 et CNRS, UMP7095, Institut d'Astrophysique de Paris, 98bis Boulevard Arago, 75014 Paris, France}
\affiliation[j]{Department of Astronomy and Astrophysics, The Pennsylvania State University, University Park, PA 16802, USA}
\affiliation[k]{Institute for Gravitation and the Cosmos, The Pennsylvania State University, University Park, PA 16802, USA}